\documentclass[graybox]{svmult}

\usepackage{hyperref}
\usepackage{url}
\usepackage{type1cm}        
\usepackage{makeidx}    
\usepackage{cite}
\usepackage{graphicx}        
\usepackage{multicol}        
\usepackage[bottom]{footmisc}

\usepackage{newtxtext}       %
\usepackage[varvw]{newtxmath}       

\makeindex             

\begin{document}

\title*{Entanglement Entropy and Causal Set Theory}
\author{Yasaman K. Yazdi}
\institute{Yasaman K. Yazdi \at Imperial College London, Blackett Laboratory, SW7 2AZ, London, UK,\newline \email{ykouchek@imperial.ac.uk}}
%
%
\maketitle

\abstract*{Each chapter should be preceded by an abstract (no more than 200 words) that summarizes the content. The abstract will appear \textit{online} at \url{www.SpringerLink.com} and be available with unrestricted access. This allows unregistered users to read the abstract as a teaser for the complete chapter.
Please use the 'starred' version of the \texttt{abstract} command for typesetting the text of the online abstracts (cf. source file of this chapter template \texttt{abstract}) and include them with the source files of your manuscript. Use the plain \texttt{abstract} command if the abstract is also to appear in the printed version of the book.}

\abstract{ We review 
a  formulation of the entanglement entropy of a quantum scalar field in terms of its spacetime two-point correlation functions. We discuss applications of this formulation to studying entanglement entropy in various settings in causal set theory. These settings include sprinklings of causal diamonds in various dimensions in flat spacetime, de Sitter spacetime, massless and massive theories, multiple disjoint regions, and nonlocal quantum field theories.}


\section{Introduction}

Entanglement entropy is a useful measure of our limited access to quantum fields' degrees of freedom. This limited access can occur for example in the presence of an event horizon, where  correlations between field values at points in the interior and exterior of the event horizon, $\langle 0|\phi(x_{int})\phi(x_{ext})|0\rangle$, are no longer available.

Entanglement entropy is one of the most important concepts in quantum gravity. It naturally combines both quantum  (entanglement) and gravitational (area laws) properties. It was originally inspired by the search for a fundamental understanding of black hole entropy: we know that black holes  classically have the Bekenstein-Hawking entropy \cite{bekenstein, hawking} associated to them, which scales as the spatial area of the event horizon, but we do not know what the fundamental or microscopic origin of this entropy is (e.g. in the statistical mechanical sense of what the microstates leading to this  entropy are). It is one of the important tasks of quantum gravity to provide insight into this. Entanglement entropy, as first shown in \cite{Sorkin:1984kjy}, also generically scales as the area of the boundary of the entangling region (which in a black hole spacetime is the area of the event horizon). Hence, it is a promising direction to investigate this question. Ultimately, we expect entanglement entropy to contribute to black hole entropy; the open question is whether or not it will be the dominant contribution.

\begin{figure}
    \centering
    \includegraphics[width=.6\linewidth]{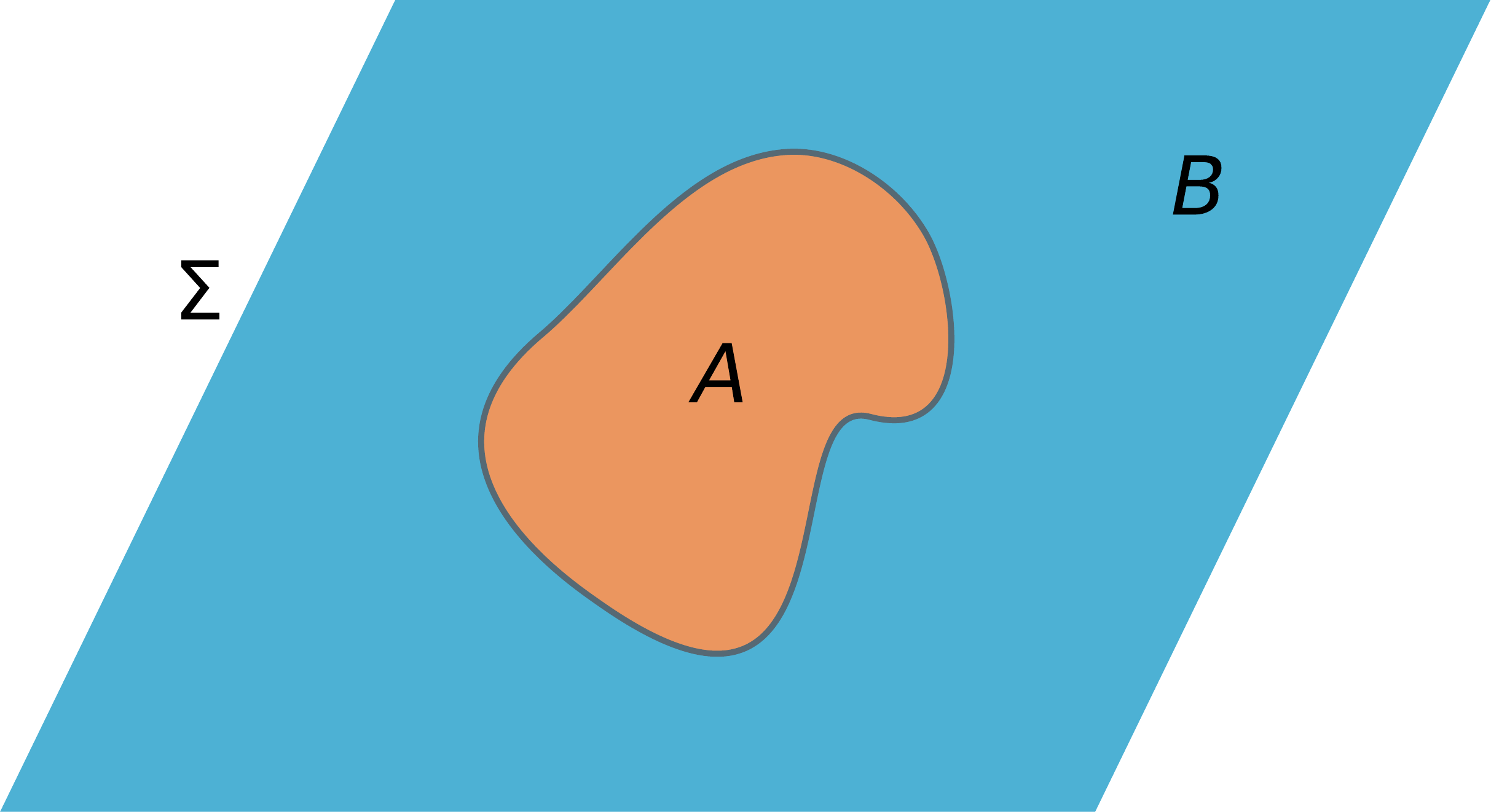}
    \caption{A spatial hypersurface $\Sigma$ divided into two complementary subregions $A$ and $B$.}
    \label{fig:Sab}
\end{figure}

Let us now review some general aspects of entanglement entropy. Entanglement entropy is conventionally defined using a density matrix $\rho$, which is initially pure, meaning that we have full information about the quantum system and the von Newumann entropy vanishes:

\begin{equation}
      S = -\text{Tr} \rho_{\text{pure}}\ln\rho_{\text{pure}}=0.
\end{equation}
Subsequently, we trace out of the density matrix the parts of the system that we do not have information about. Traditionally this tracing out is done on a spatial hypersurface $\Sigma$, as in Fig. \ref{fig:Sab}, which is divided into two complementary subregions, region $A$ and region $B$, one of which represents the degrees of freedom we do have access to and the other the degrees of freedom that we do not have access to. After we trace out the degrees of freedom in one of the subregions, for example those in $B$\footnote{We would get the same answer if we instead traced out the degrees of freedom in $A$ to get $\rho_B$ and computed $S_B$. We refer to this as complementarity below.}, we get a reduced density matrix

\begin{equation}
    \rho_A = \text{Tr}_B \rho,
    \end{equation}
and the entropy of entanglement between region $A$ and $B$ is defined as \cite{Sorkin:1984kjy}

\begin{equation}
   S_A = -\text{Tr}\rho_A\ln\rho_A  \ .
   \label{ee}
\end{equation}

It is crucial that the theory one is working with has a UV cutoff with respect to which we count how many degrees of freedom we do or do not have access to. Without a UV cutoff, we would get an infinite answer for the entanglement entropy in \eqref{ee}.

Since the original work \cite{Sorkin:1984kjy} in the context of black hole entropy, entanglement entropy has found many additional useful applications in other areas of physics such as quantum information \cite{Wootters:1997id} and condensed matter physics \cite{Laflorencie:2015eck}. Depending on the specific application in mind, different techniques may be used to evaluate the entropy in \eqref{ee}. This choice of technique is often motivated by physical and computational considerations.

\begin{figure}
    \centering
    \includegraphics[width=.45\linewidth]{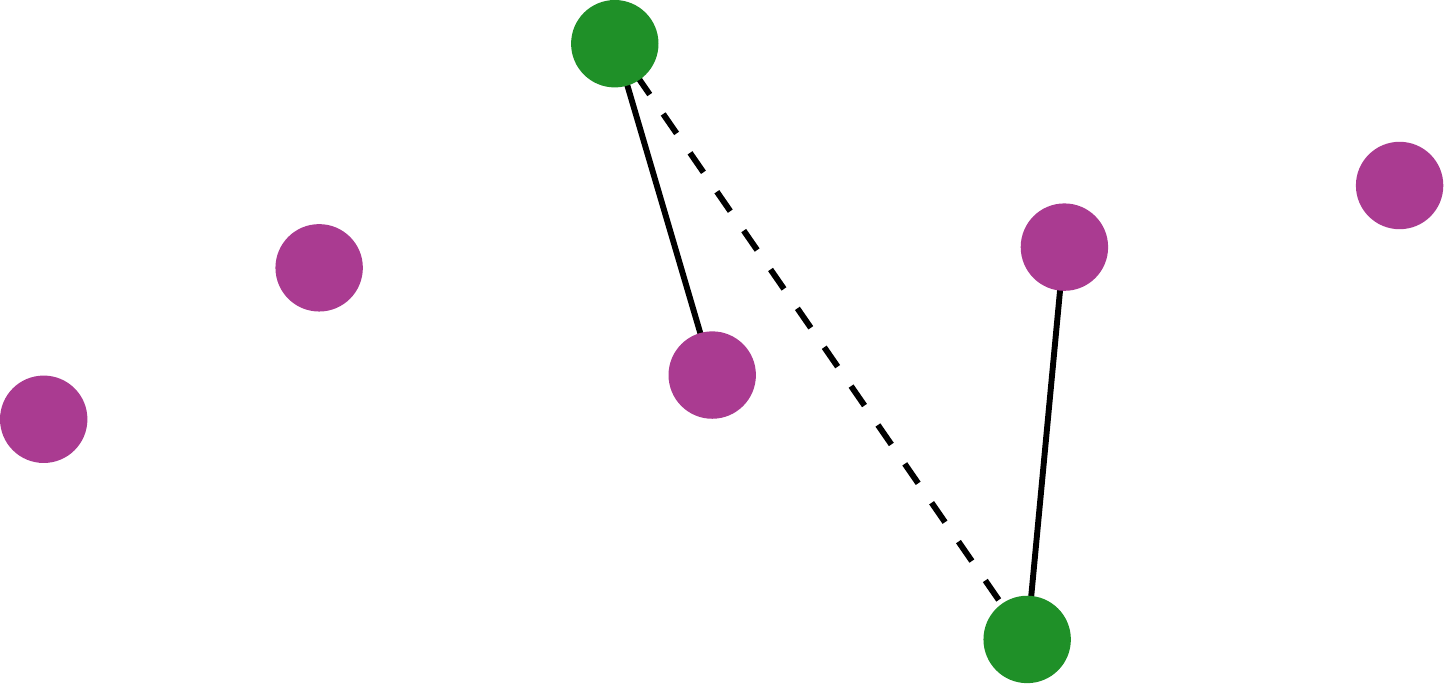}
    \caption{The pink elements form a maximal antichain, which is the largest set of unrelated elements such that every other element is to the past or future of at least one of the elements of this set. This is the analogue of a spatial hypersurface in a causal set.  In order for it to be a viable analogue of a Cauchy surface, we must be able to deduce all causal relations using relations involving the antichain elements. However, this is not possible, as illustrated by the  counterexample above: the causal relation represented by the dashed line has no imprint on the antichain, hence we do not have an analogue of a Cauchy hypersurface in a causal set.}
    \label{fig:sieve}
\end{figure}

In causal sets, there is no analogue of field data on a spatial hypersurface (i.e. a Cauchy surface). Fig. \ref{fig:sieve} illustrates the reason for this, which is essentially that there is no guarantee that there will not be relations that will not make an imprint on a subset of unrelated elements. Therefore we cannot work on a hypersuface as in Fig. \ref{fig:Sab} and must use an intrinsically spacetime approach to compute the entanglement entropy. Fortunately, a spacetime definition of entanglement entropy, in terms of the spacetime two-point correlation function, exists and can be used in causal set calculations. We review this formulation in Section \ref{sec:see}. While we are led to a spacetime formulation of entanglement entropy in causal set theory out of necessity, there are in fact numerous attractive features of working with a spacetime definition of entanglement entropy. For example, quantum fields themselves are really spacetime quantities (their domain is spacetime). Therefore, it is more natural to study them in spacetime. Furthermore, quantum fields are highly singular and may not always admit meaningful restrictions to spatial hypersurfaces. We devote Section \ref{sec:QFST} to supporting these statements with some concrete examples. Finally, in Section \ref{sec:apps} we discuss several calculations of entanglement entropy, in settings including sprinklings into flat spacetime and de Sitter spacetime, before ending in Section \ref{sec:end} with a discussion of some subtleties of and the future directions for this work.

\section{Entanglement Entropy from Spacetime Two-point Correlation Functions}
\label{sec:see}
A quantum field theory with local operators is typically fully determined by the set of all its n-point correlation functions. If we consider a scalar field $\phi$, these are $\langle 0|\phi(x_1) \phi(x_2)|0\rangle$,... $ \langle 0|\phi(x_1) \phi(x_2)\dots\phi(x_n)|0\rangle$, where $\{x_1,x_2,...\}$ represent points in spacetime. In a Gaussian quantum field theory, things are much simpler because we only need to know the two-point correlation function $\langle 0|\phi(x_1) \phi(x_2)|0\rangle$. In this case, all the higher n-point correlation functions are derivable from the two-point function through Wick's theorem. For the remainder of this chapter, except where explicitly mentioned otherwise, we will restrict the discussion to  Gaussian scalar field theories.

\subsection{Entropy from Correlation Functions}
\label{subsec:entropy}
Since we are focusing on a Gaussian scalar field theory, as mentioned above, everything (including the entanglement entropy \eqref{ee}) must be expressible in terms of the two-point correlation function $\langle 0|\phi(x) \phi(x')|0\rangle$. The definition of entropy introduced in \cite{Sorkin:2012sn}, which we review in this subsection, does precisely that. Specifically, the entropy is given by the sum 

\begin{equation}
S=\sum_{\lambda} \lambda\ln |\lambda|
\label{s4}
\end{equation}
over the solutions $\lambda$ to the generalized eigenvalue problem 

\begin{equation}
 W \, v = i\lambda \; \Delta \, v,  \            
\label{geneig}
\end{equation}
while excluding components in the kernel of $\Delta$

\begin{equation}
\Delta v\neq 0 ~.
\label{kernel}
\end{equation}
$W$ in \eqref{geneig} is the  spacetime two-point correlation function or Wightman function, $W(x,x')=\langle 0|\phi(x) \phi(x')|0\rangle$, and $i\Delta$ is the Pauli-Jordan function or spacetime commutator of the field, $i\Delta(x,x')=[\phi(x), \phi(x')]$. In a Gaussian theory, $i\Delta$ is a c-number. We can obtain $\Delta$ using the retarded Green function, $G_R$, as $\Delta(x,x')=G_R(x,x')-G_R(x',x)$. If we already have a state and therefore a $W$ to work with, we can also obtain it through the imaginary or anti-symmetric part of $W$, i.e. $i\Delta(x,x')=W(x,x')-W(x',x)=2~\text{Im}(W(x,x'))$.

\begin{figure}
    \centering
    \includegraphics[width=.6\linewidth]{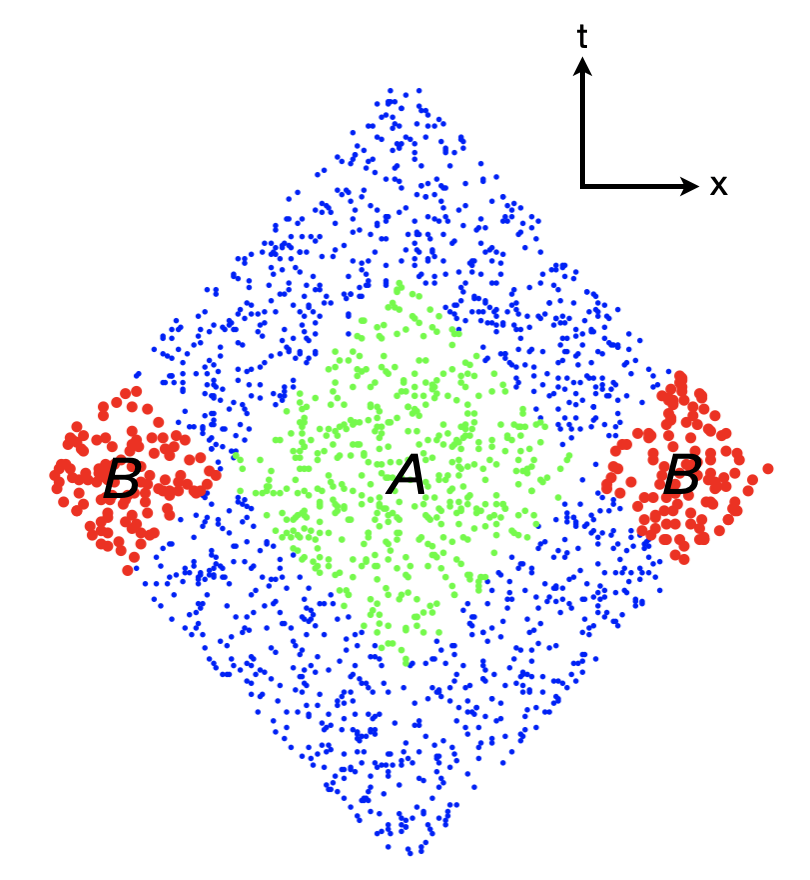}
    \caption{Causally complementary spacetime subregions $A$ and $B$ in a causal set.}
    \label{fig:complem}
\end{figure}

More precisely, we must start with a global state (and its corresponding $W$) which is initially pure. In terms of the eigenvalue equation \eqref{geneig} and entropy \eqref{s4}, purity would mean solutions $\lambda$ that are $1$'s and $0$'s. In the next subsection we review one choice of pure state, the Sorkin-Johnston state, that can be defined in causal set theory. With a pure state at hand, we subsequently exclude parts of the quantum system that we don’t have access to by restricting the elements $x$ and $x’$ in $W(x,x')$ and $i\Delta(x,x')$ to lie in the \emph{spacetime} subregion that we do have access to, before solving \eqref{geneig}. Then we can interpret the resulting entropy from substituting the solutions into \eqref{s4} as the entanglement entropy between the spacetime region which is the domain of $x$ and $x'$ and its causal complement. Note that the causal complement will not in general be the complementary spacetime domain (see e.g. Fig. \ref{fig:complem}).

It is in this way, owing to  $W(x,x')$ and $i\Delta(x,x')$ being spacetime functions, that this formulation of entanglement entropy is a spacetime approach. Additionally, due to its spacetime nature it allows one to use a spacetime UV cutoff such as the discreteness scale of a causal set. This can give our counting of degrees of freedom a covariance and universality that is not present in the more spatial formulations. 

We can in some cases compare the results obtained from this method to the results from the conventional spatial methods. We can do this for example when the regions of spacetime that we consider are domains of dependence of Cauchy surfaces. Fig. \ref{fig:complem} illustrates an example of this, since the causal diamonds are domains of dependence of the $1$D intervals connecting the left and right spatial corners (e.g. their diameters).  

This entropy formulation is derived in \cite{Sorkin:2012sn}. An alternative derivation of it using the replica trick can be found in \cite{Chen:2020ild}. Quite surprisingly, this formula can actually be used beyond Gaussian theories. In \cite{Chen:2020ild} it was shown that up to first order in perturbation theory, the entanglement entropy of even  non-Gaussian or interacting theories, is captured by the two-point correlation function via  \eqref{geneig} and \eqref{s4}. The only difference in the interacting theory case is that the $W$ that enters \eqref{geneig} is the interacting one and while $i\Delta$ would still be the antisymmetric part of $W$, it is no longer the commutator.

\subsection{The Sorkin-Johnston (SJ) Vacuum}

The Sorkin-Johnston (SJ) Wightman function \cite{sj1, sj2} is defined in the same algebraic and spacetime spirit as the entropy formulation we have just reviewed. It uniquely picks out a vacuum state in any globally hyperbolic spacetime. 

 To define the SJ state, we require the spacetime commutator function $i\Delta$. As we saw above, we can express $i\Delta$ in terms of the retarded Green function\footnote{$
(\Box+m^{2})G_R(x,x')=-\frac{\delta^{D}(x-x')}{\sqrt[]{-g}}$, 
where $G_R(x,x')$ is only non-zero if $x'\prec x$.} as
 \begin{equation}
 \Delta(x,x')=G_R(x,x')-G_R(x',x) .   
 \end{equation}
  $i \Delta$ is anti-symmetric and Hermitian.  We can then rewrite it as an expansion in terms of its positive and negative eigenvalues and their respective eigenfunctions as
  \begin{equation}
i\Delta=\sum_i\tilde{\lambda}_iu_iu^\dagger_i-\tilde{\lambda}_iv_iv^\dagger_i,
\label{delta}
\end{equation}
  where $u_i$ and $v_i$ are  the normalized positive and negative eigenvectors respectively, and $\tilde{\lambda_i}>0$.  The SJ Wightman function is defined by restricting to the positive eigenspace of $i\Delta$
  
  \begin{equation}
    W_{SJ}:=\text{Pos}(i\Delta)=\sum_i\tilde{\lambda}_iu_iu^\dagger_i.
    \label{w}
\end{equation}

  When we define W in this way, the only solutions to the generalized eigenvalue equation \eqref{geneig} are $\lambda=0$ or $\lambda=1$, and this makes the entropy \eqref{s4} vanish, as required if the state is pure. It is also worth mentioning that in static spacetimes, the SJ vacuum is the same one that is picked out by the timelike and hypersurface-orthogonal Killing vector \cite{Aslanbeigi}, hence the SJ state is known to reflect the symmetries of the background geometry if there are any. In all of the entanglement entropy applications that we will review in Section \ref{sec:apps}, the SJ state is used as the pure state.

Both the entanglement entropy formulation and SJ state prescription can also be used in continuum spacetimes. Before moving on to their applications in causal set theory, we will in the next section motivate why it is desirable to work with spacetime formulations, in general, when studying quantum field theories.

\section{Quantum Field Operators are Distributions in Spacetime}
\label{sec:QFST}

At least as early as a 1933 work by Bohr and Rosenfeld \cite{bohr}, it has been recognized that quantum field operators are only well-defined as averages over finite regions of spacetime rather than at individual spacetime points. This averaging is achieved through smearing the quantum field with smooth, real-valued functions with compact support

\begin{equation}
    \phi(f)=\int \phi(x) f(x)\, dV.
\end{equation}
 See also \cite{geroch} for a modern review of the convergence issues that arise if quantum fields are defined at points in spacetime rather than as distributions.
Less well-known is the fact that generic nonlinear operators in free quantum field theories are not well-defined if they are smeared with  functions with support only on spatial hypersurfaces rather than spacetime regions  \cite{streater, haag}. They become well-defined  only after some smearing in a duration of time as well. Furthermore, based on perturbation theory, then, it is expected that operators in generic interacting quantum field theories are also not well-defined on spatial hypersurfaces, since they would contain the ill-defined nonlinear terms from the free theory. 

Below we study this behaviour for two generic nonlinear operators ($\phi^2$ and $T_{ab}$) in free scalar field theory  in Minkowski spacetime. 

\subsection{Variance of a Free Scalar Field $\phi^2$ Operator}

Consider the usual quantum scalar field operator in $3+1$ spacetime dimensions 

\begin{equation}
    \phi(x)=\int\frac{d^3p}{(2\pi)^3\sqrt{2 E_{\vec{p}}}}\left(a_{\vec{p}}\, e^{ip\cdot x}+a^\dagger_{\vec{p}}\, e^{-ip\cdot x}\right),\end{equation} where $E_{\vec{p}}=\sqrt{|\vec{p}|^2+m^2}$ and $x=(t,\vec{x})$. 
   
Let us now examine the normal-ordered $\phi^2$ operator, 
and smear it with a test function $f(x)=f_1(t) f_2(\vec{x})$ with compact support on a time  $t'=\text{const}$ slice, i.e. $f_1(t)=\delta(t-t')$, 
\begin{equation}
:\phi^2(t', f):\,=\int dt\, d^3x\, f_2(\vec{x})\delta(t-t')    \int \frac{d^3p \, d^3k}{2 (2\pi)^6 \sqrt{E_{\vec{p}}E_{\vec{k}}}}\left(a_{\vec{p}}\,a_{\vec{k}}e^{i(p+k)\cdot x}+\dots\right).\,\,\,\,
\end{equation}
We then square the result 
and compute its expectation value in the Minkowski vacuum,
\begin{align}
  \langle& 0|  :\phi^2(t', f_2)::\phi^2(t', f_2):|0\rangle=\int \frac{d^3p \, d^3k |\tilde{f}_2(\vec{k}+\vec{p})|^2}{2 (2\pi)^{12} E_{\vec{p}}E_{\vec{k}}},
  \label{sqnoe}
\end{align}
where $\tilde{f}_2$ is the Fourier transform of $f_2$.  The integral can be seen to be linearly divergent by simply counting the powers of $\vec{p}$ and $\vec{k}$, and assuming that $f_2$ is square-integrable.
However, \eqref{sqnoe} converges if the smearing is done over an extent in time as well. To see this, let us for simplicity assume that the smearing function is separable, i.e., $f(x)=f_1(t) f_2(\vec{x})$ as before (where $f_1$ is no longer a delta function) and both $f_1$ and $f_2$ are square-integrable. Then
 \begin{align}
 \langle 0|  :\phi^2(f)::\phi^2( f):|0\rangle
 &\propto\int \frac{d^3p\, d^3k\, |\tilde{f}_2(\vec{p}+\vec{k})|^2\,|\tilde{f}_1(p^0+k^0)|^2}{p^0 k^0} \label{stsmear}\\
 &\leq \int \frac{d^3k\,  I(\vec{k}) \tilde{F}_1(k^0)^2}{\sqrt{|\vec{k}|^2+m^2}},
 \end{align}
 with $\tilde{F}_1(k^0)$ defined as the absolute maximum of $\tilde{f}_1$ with respect to $p^0$ at a given $k^0$, and 
 $I(\vec{k})$ being the $p$-integral of the remaining $p$-dependent variables.  Since it can be shown that  $I(\vec{k})\sim\frac{1}{|\vec{k}|}$ for large $|\vec{k}|$, and $\tilde{F}_1$ is square-integrable, we see by counting powers of $k$ that the integral is convergent.

 Expectation values such as the one we have considered above are ubiquitous in operator product expansions (OPEs) and appear in generic correlation functions. Similar divergences occur for other nonlinear operators such as $\phi^n$ for $n\geq2$ and the stress energy tensor $T_{ab}$ (as we show below), and suggest that quantum field theories generally need to be considered in a spacetime region.

\subsection{Variance of Scalar Field Energy Density $T_{00}$}

The simple construction above shows that certain operator expectation values are divergent when smeared only on a spatial hypersurface and convergent when smeared in spacetime. 
Let us next consider the analogous calculation for a physically more interesting quantity, namely, the scalar field's stress-energy tensor
\begin{equation}
    :T_{ab}:=:\partial_a\phi\partial_b\phi:-\frac{1}{2}\eta_{ab}:\partial_c\phi\,\partial^c\phi:-\frac{m^2}{2}\eta_{ab}:\phi^2:
\end{equation}
We will focus our attention on the energy density component $:T_{00}:$, and show that its variance diverges when smeared on a hypersurface, but can be made convergent with an appropriate spacetime smearing.  Substituting in the definition of $\phi$ and smearing with a square integrable test function $f(x)=\delta(t-t') f_2(\vec{x})$ defined on a timeslice, the expectation value of the variance is given by
\begin{align}
  \langle 0|& :T_{00}(f):^2\,|0\rangle=\int \frac{d^3p \, d^3k }{8 (2\pi)^{12}}\left(\frac{E_{\vec{p}}E_{\vec{k}}+\vec{p}\cdot\vec{k}-m^2}{\sqrt{E_{\vec{p}}}\sqrt{E_{\vec{k}}}}\right)^2|\tilde{f}_2(\vec{p}+\vec{k})|^2.
  \label{t001}
\end{align}
We then see in the large $|\vec{p}\,|$ and $|\vec{k}\,|$ limit that the leading order contribution to the integral is
\begin{align}
  \langle 0|& :T_{00}(f):^2\,|0\rangle\sim\int d^3p \, d^3k\,|\vec{p}\,|\, |\vec{k}\,|\left(1+\cos\theta\right)^2|\tilde{f}_2(\vec{p}+\vec{k})|^2,
  \label{t002}
\end{align}
$\theta$ being the angle between $\vec{p}$ and $\vec{k}$ in the dot product above.  Since this integral has higher powers of $\vec{p}$ and $\vec{k}$ than the integral in \eqref{sqnoe} which we already showed diverges, it follows trivially that \eqref{t002} diverges as well.

In order for a time smearing to make \eqref{t002} converge, more care needs to be taken than in \eqref{stsmear} where generic square integrable smearing functions suffice. Here we have in effect four extra factors of  $\vec{p}$ and $\vec{k}$, so we need to ensure that the Fourier transforms of our smearing functions decay fast enough near infinity to counteract this.  For this reason and for simplicity of analysis, we will use Gaussian smearing functions.\footnote{While technically a Gaussian does not have compact support, its rapid decay restricts meaningful values to a local enough region.  In any case, one could also work with ``bump'' functions which have compact support, but nonetheless have Fourier transforms that decay faster than any power law, as demonstrated in \cite{bump}.} We again write the spacetime smearing function as the separable function $f(t, \vec{x}) = f_1(t)\,f_2(\vec{x})$, where
\begin{align}
    f_1(t) = e^{-a \, t^2} &\implies |\tilde{f}_1(p^0 + k^0)|^2 \propto e^{-\frac{\left(p^0 + k^0\right)^2}{2\,a}},\\
    f_2(\vec{x}) = e^{-b \, \vec{x}\,^2} &\implies |\tilde{f}_2(\vec{p} + \vec{k})|^2 \propto e^{-\frac{\left(\vec{p} + \vec{k}\right)^2}{2\,b}}.
\end{align}
The variance of the spacetime smeared energy density then at large $|\vec{p}\,|$ and $|\vec{k}\,|$ has leading order contribution
\begin{align}
  \langle 0|\,\,:T_{00}(f):^2\,|0\rangle&\sim\int d^3p \, d^3k\,|\vec{p}\,|\, |\vec{k}\,|\left(1+\cos\theta\right)^2\,|\tilde{f}_1(p^0 + k^0)|^2\,|\tilde{f}_2(\vec{p}+\vec{k})|^2\\
  &\lesssim \int d^3k \,|\vec{k}\,|\,\tilde{f}_1(k^0)|^2\,I(\vec{k}), \label{t003}
\end{align}
where the inequality follows from $\tilde{f}_1(v+w) < \tilde{f}_1(v)$ for all positive $v$, $w$ when $\tilde{f}_1$ is a Gaussian, and we have simply grouped the remaining terms into the $p$ integral $I$.  One can then asymptotically evaluate $I$ as
\begin{align}
    I(\vec{k}) &\propto \int d^3p \,|\vec{p}\,|\left(1+\cos\theta_p\right)^2e^{-\frac{\left(\vec{p} + \vec{k}\right)^2}{2\,b}} \\
    &\sim \mathcal{O}(|\vec{k}\,|^4 \log(|\vec{k}\,|))\,+\,\mathcal{O}(|\vec{k}\,|)\,+\,\mathcal{O}(|\vec{k}\,|^2e^{-\frac{\vec{k}\,^2}{2\,b}}),
\end{align}
in the large $|\vec{k}\,|$ limit.  Replacing the leading-order term back into \eqref{t003}, the upper bound on the variance of the energy density becomes asymptotic to the $|\vec{k}\,|$-integral
\begin{align}
  \langle 0|\,\,:T_{00}(f):^2\,|0\rangle&\lesssim \int^\infty_0 d|\vec{k}\,| \,|\vec{k}\,|^7 \log(|\vec{k}\,|)\,e^{-\frac{|\vec{k}\,|\,^2}{2\,a}}.
\end{align}
The decaying Gaussian will dominate asymptotically, so this integral is finite. 

Therefore, as anticipated, $\langle 0| :T_{ab}::T_{ab}:|0\rangle$ is  well-defined only as a distribution in spacetime. The smearing in the above discussion can be considered as a model for making a measurement. We would certainly not expect the energy in a bounded region to have an infinite
variance, but we see that a finite variance is only obtained if the bounded region is in spacetime rather than in space alone. Hence we can conclude that the quantum stress tensor is only meaningful as a distribution in \emph{spacetime}. See also \cite{ford} for evidence that a well-defined probability distribution for the quantum stress tensor is only obtained for averages in time or spacetime. 

\section{Applications}
\label{sec:apps}
Having established the importance of entanglement entropy in quantum gravity, as well as the importance of treating quantum fields in a spacetime domain, let us now explore what work on entanglement entropy in causal set theory has been done.

As a reminder, the calculation of entanglement entropy schematically amounts to 

\begin{equation}
    G_R\rightarrow i\Delta \rightarrow W_{SJ} \rightarrow S.
    \label{scheme}
\end{equation}

However, note that the scheme above does not tell one how to obtain the retarded Green function, $G_R$, which is the starting point. In fact, in general there exists no known expression for $G_R$ in terms of quantities intrinsic to the causal set (such as the link or causal matrices). The examples below represent some important and interesting cases where an expression for $G_R$ is known. 

\subsection{Causal Diamond in $1+1$D Flat Spacetime}
The most studied and well-understood setting for entanglement entropy in a causal set, is the causal diamond in $1+1$D. This setting also benefits from numerous analytic results from the continuum being known and available for comparison. 

We will mainly focus on the massless theory, for simplicity. The retarded Green function in $1+1$D Minkowski spacetime, for a massless scalar field theory is 

\begin{equation}
    G_R(x,x')=\frac 1 2\theta(t-t')\theta(\tau^2),
\end{equation}
where $\tau=\sqrt{|t-t|^2-|\vec{x}-\vec{x'}|^2}$ is the proper time between the two points and $\theta$ is the Heaviside step function. In other words, $G_R$ is only nonzero if $x'$ causally precedes $x$ and when it does, $G_R$ takes the value $\frac 1 2$. With $G_R$ so directly related to the causal structure, we have an exact analogue of it in the causal set in terms of the causal matrix $C$. The causal matrix similarly has a nonzero value of $1$ for each entry $C_{x'x}$ where $x'$ causally precedes $x$ or $x'\prec x$. Therefore, all we have to do is multiply the causal matrix by the constant $\frac 1 2$ to get the analogue of the retarded Green function in the causal set:

\begin{equation}
    G_R=\frac 1 2 C
\end{equation}

Next, we form $i\Delta$ from $\frac i 2$ times the antisymmetric part of the causal matrix\footnote{When we write $C_{xy}$, we mean the entry corresponding to elements $x$ and $y$ in a point basis representation of the the matrix $C$.}

\begin{equation}
    i\Delta_{xx'}=\frac i 2 (C_{x'x}-C_{xx'}).
    \label{idelta2d}
\end{equation}

From here  it is a simple algebraic exercise to diagonalize \eqref{idelta2d} and restrict to its positive eigenspace to define $W_{SJ}$. The SJ state in the $1+1$D causal diamond has been extensively studied \cite{Afshordi_2012}. It resembles the standard Minkowski Wightman function (with an IR cutoff) away from the boundaries of the diamond. Near the left and right corners it resembles the Minkowski Wightman function with a static mirror at these corners. 
\begin{figure}
    \centering
    \includegraphics[width=.75\linewidth]{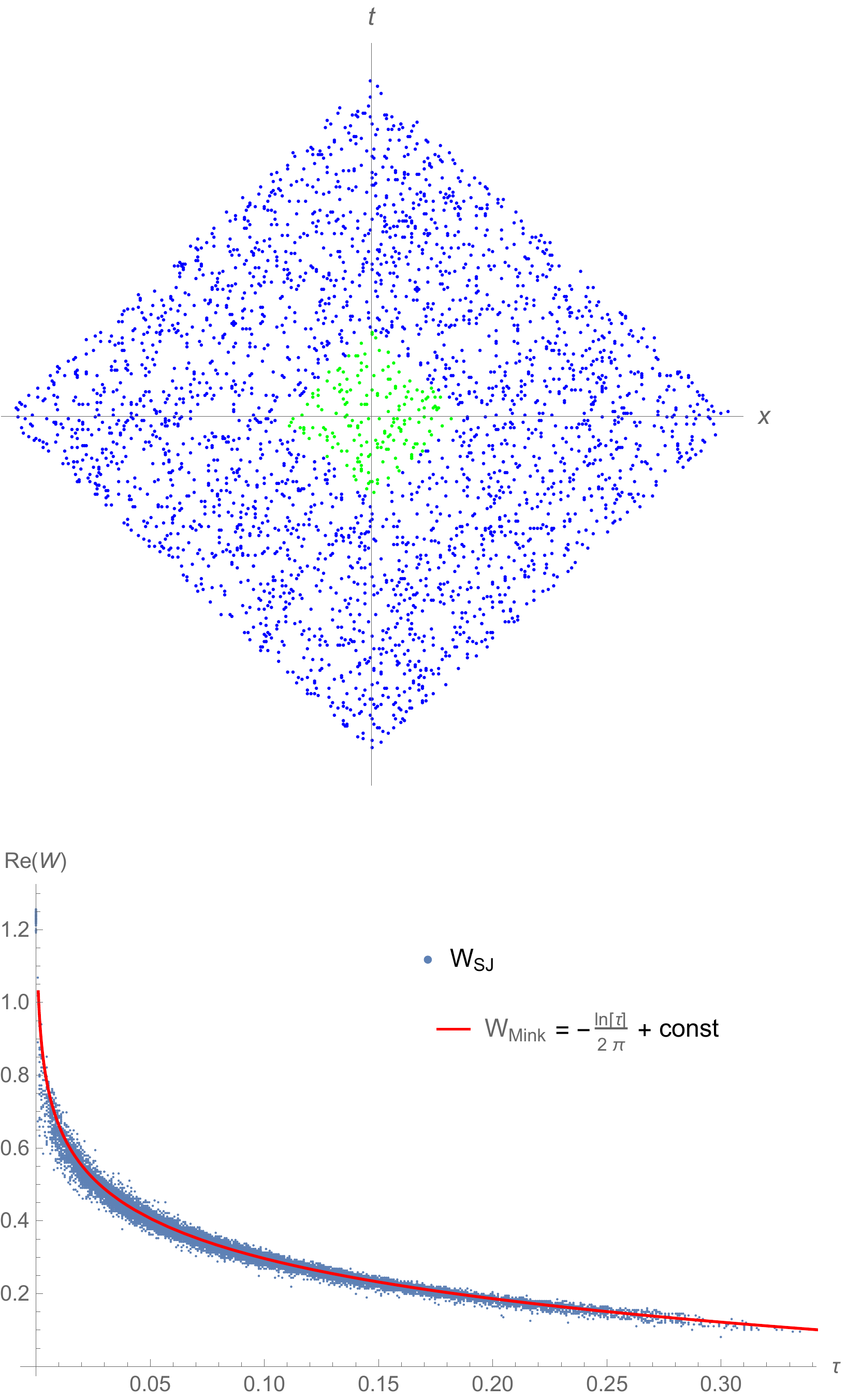}
    \caption{The lower plot shows the SJ Wightman function in a causal set causal diamond, away from the boundaries. The SJ state is computed in the larger blue diamond at the top. The values of $W_{SJ}$ versus proper time have been shown for elements in the inner Green subdiamond. Agreement is seen with the Wightman function associated with the IR-regulated Minkowski vacuum state (red curve).}
    \label{fig:sjdiamond}
\end{figure}

In \cite{Sorkin_2018, Yazdi:2017pbo} it was shown that if we go ahead and compute the entanglement entropy according to the steps in Section \ref{subsec:entropy} , we obtain an unexpected answer: instead of the usual spatial area law scaling of the entanglement entropy with the UV cutoff, we obtain a \emph{spacetime volume} scaling. Since we have $N\propto V$ in a causal set, a volume scaling means that the entanglement entropy scales linearly with $N$ instead of the expected $N^{D-2}$ scaling in $D>2$ or logarithmic scaling when $D=2$. 

\begin{figure}
    \centering
    \includegraphics[width=.7\linewidth]{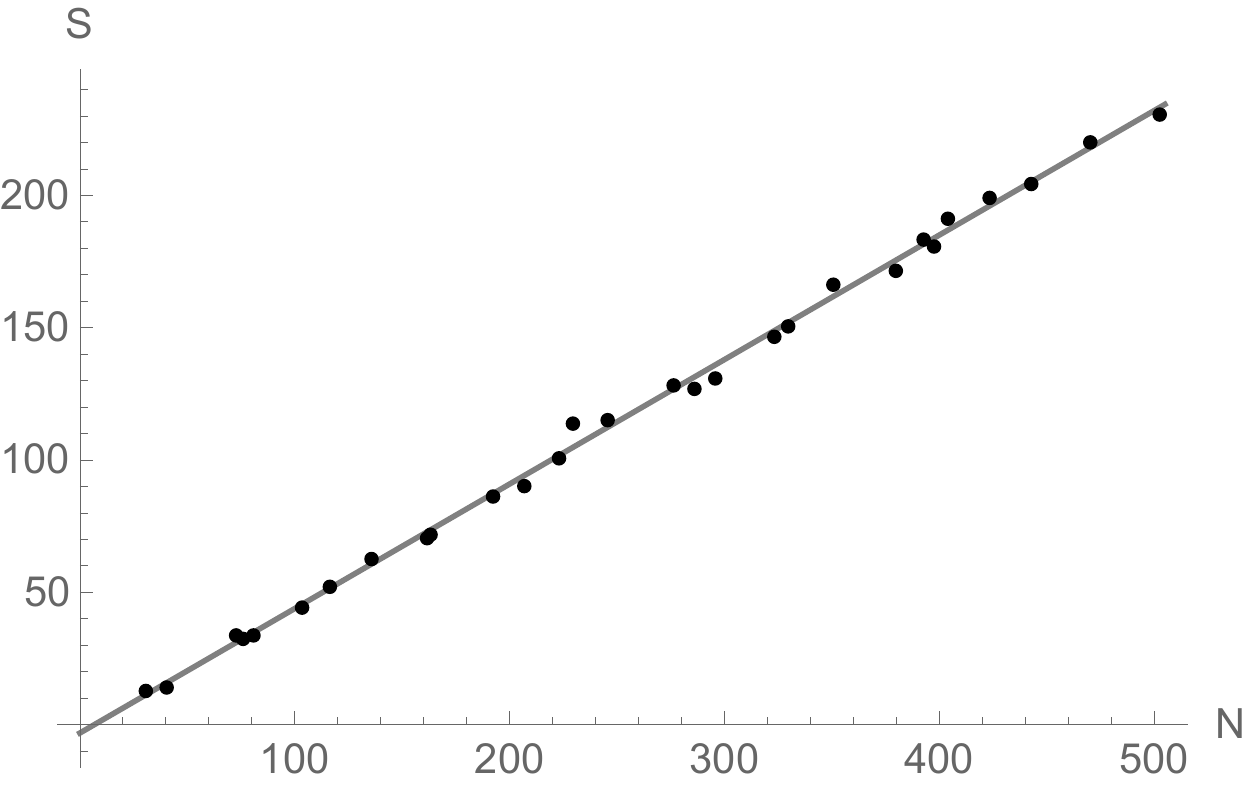}
    \caption{The entanglement entropy grows linearly with $N$, demonstrating a spacetime volume law scaling. The points represent entropy values from the  calculation of \eqref{s4} in the setup of the upper plot in Fig.\ref{fig:sjdiamond}. The line is a linear fit to the data. In this example, the ratio of the side lengths of the two diamonds was $\frac 1 4$.}
    \label{fig:Slin}
\end{figure}

More specifically, to study the scaling of the entanglement entropy with the UV cutoff, which in length dimensions is $\rho^{-1/D}$ (where $\rho$ is now the sprinkling density), we fix the geometry into which the sprinkling is performed (e.g. the blue diamond in Fig. \ref{fig:sjdiamond}), as well as the entangling subregion (e.g. the green subdiamond in Fig. \ref{fig:sjdiamond}). We then vary the number of elements $N$ that we sprinkle, thereby varying $\rho$ and the UV cutoff. Fig. \ref{fig:Slin} shows an example of the volume law scaling obtained, for the diamond setup shown in Fig. \ref{fig:sjdiamond}.

This result is peculiar to the causal set calculation, as the continuum analogue of the same calculation (performed in \cite{Saravani_2014}), showed no sign of this behaviour. A closer look at the eigenvalues of $i\Delta$, with the help of insight from analytic results from the continuum, reveals the source of the extra entropy. In the continuum diamond with side length $2\ell$, the eigenfunctions of $i\Delta$ with nonzero eigenvalues are \cite{SJthesis}

\begin{align}
  f_k(u,v) &= e^{-iku} - e^{-i k v}, & &\textrm{with } k = \frac{n \pi}{{\ell}}, \; n =\pm 1,\pm 2, \ldots\label{eq:SJfunctions1}\\
  g_k(u,v) &= e^{-iku} + e^{-i k v} - 2 \cos(k {\ell}), & &\textrm{with } k|\tan(k{\ell})=2k{\ell} \wedge k\in\mathbb{R}\,\label{eq:SJfunctions2} \end{align}

The eigenfunctions above have been expressed in terms of lightcone coordinates $u=\frac{t+x}{\sqrt{2}}$ and $v=\frac{t-x}{\sqrt{2}}$. The eigenvalues are ${\tilde{\lambda}}_k={\ell}/k$, with  eigenvalues from both sets of eigenfunctions $f_k$ and $g_k$ approaching  ${\tilde{\lambda}}_k=\frac{\ell^2}{n\pi}$ in the large $k$ limit. The eigenfunctions \eqref{eq:SJfunctions1} and \eqref{eq:SJfunctions2} span the solutions of the Klein Gordon equation.\footnote{$\text{Ker}(\Box+m^2)=\overline{\text{Im}(\Delta)}$.} Therefore, if we were to consider a finite number of eigenfunctions up to some $k_{max}$, we can think of $k_{max}$ as a cutoff. In other words, with a finite set of eigenfunctions up to $k_{max}$, we would only be able to expand solutions (e.g. initial data) up to this maximum wavenumber. Turning this argument around, if there were reason to believe that solutions beyond some UV cutoff $k_{max}$ ought not to be supported in a setting, we would expect to obtain a finite number $\sim n_{max}$ of eigenvalues and eigenfunctions corresponding to $k_{max}=\frac{n_{max}\pi}{\ell}$. This is precisely the scenario we are faced with in  causal sets. In our $1+1$D diamond causal set, the discreteness length is $\frac{1}{\sqrt{\rho}}=\sqrt{\frac V N}=\frac{2\ell}{\sqrt{N}}$. Hence we do not expect to be able to meaningfully describe wavelengths shorter than this. Converting this wavelength to a wavenumber we get $\frac{2\ell}{\sqrt{N}}=\frac{2\pi}{k_{max}}\implies k_{max}=\frac{\pi \sqrt{N}}{\ell}\implies n_{max}\approx 2 \sqrt{N}$, where the factor of $2$ in $2 \sqrt{N}$ comes from the fact that we have two sets of eigenfunctions $f_k$ and $g_k$ that will each have this $k_{max}$ and contribute $\sqrt{N}$.

\begin{figure}
    \centering
    \includegraphics[width=.7\linewidth]{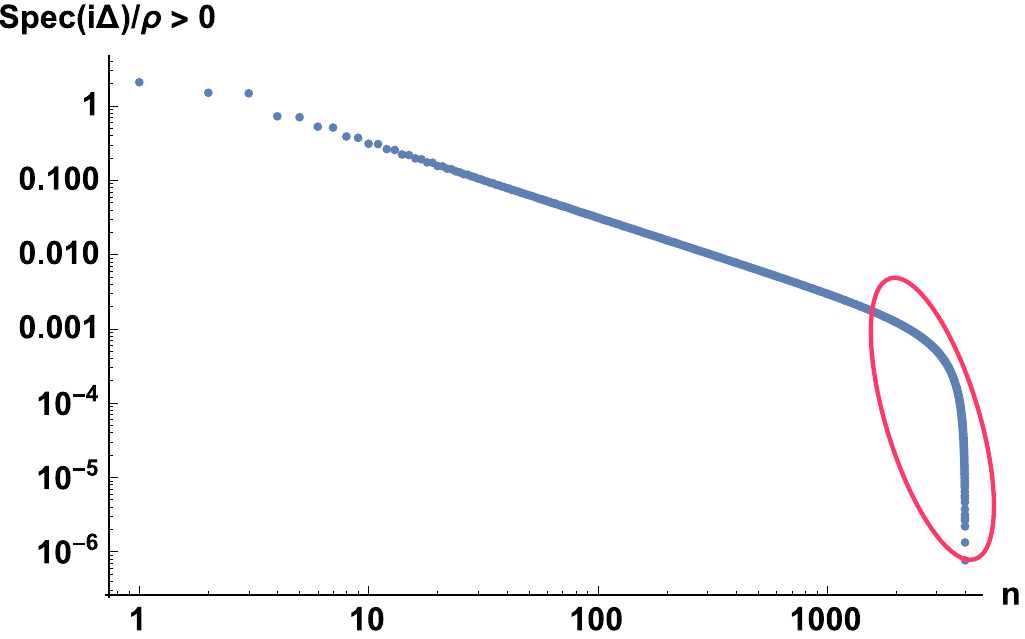}
    \caption{The positive eigenvalues of $i\Delta$ for a sample  sprinkling into a causal diamond.}
    \label{fig:eigs}
\end{figure}

However, a look at the eigenvalues and eigenfunctions of $i\Delta$ in the causal set reveals that we in fact end up with many more nonzero eigenvalues than $n_{max}$. An example of the spectrum of $i\Delta$ is shown in Fig. \ref{fig:eigs}. The values of the positive eigenvalues are shown on a log-log scale, where they have been ordered from largest to smallest, and each $i^{th}$ eigenvalue in this ordering is paired with $i$ on the horizontal axis. Circled are the extra eigenvalues that are beyond the expected $n_{max}$. Interestingly, these extra eigenvalues behave qualitatively differently from the rest: they do not follow a power law like the larger eigenvalues. 

\begin{figure}
    \centering
    \includegraphics[width=.7\linewidth]{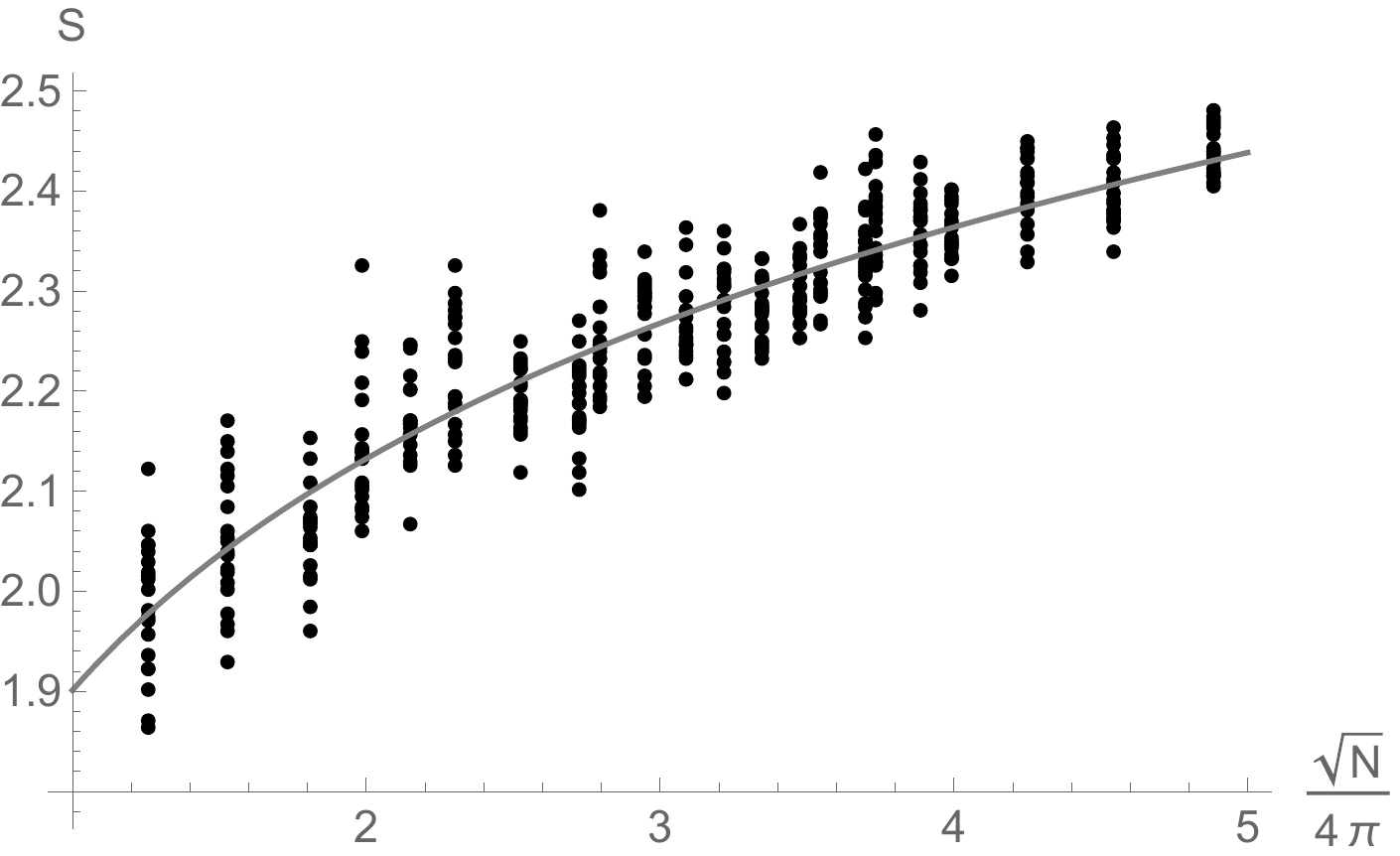}
    \caption{The entanglement entropy versus the UV cutoff, following a logarithmic scaling with a coefficient consistent with $\frac 1 3$. This is the conventional result according to the expectation of a spatial area law.}
    \label{fig:arealaw}
\end{figure}

When the contributions to the entanglement entropy from these extra eigenvalues and their corresponding eigenvectors are removed, we recover the expected spatial area law scaling of the entanglement entropy with respect to the UV cutoff. This removal is referred to as a \emph{truncation}, and it must be implemented at two stages of the calculation: (1) A first truncation when $W_{SJ}=\text{Pos}(i\Delta)$ is constructed, and (2) a second truncation when the generalized eigenvalue equation \eqref{geneig} is solved. These truncations can also be regarded as projections down to the subspace where $k_{max}=\frac{\pi \sqrt{N_\diamond}}{\ell_\diamond}$, where the diamond subscript indicates that in the first truncation which is in the global diamond, $N$ and $\ell$ are the total number of elements and size of this diamond, whereas in the second truncation which occurs in the smaller subdiamond, $N$ and $\ell$ are the number of elements in and size of the subdiamond.  When this so-called ``double truncation" is performed, we obtain the result shown in Fig. \ref{fig:arealaw}

Note that, as mentioned earlier, a logarithmic scaling with respect to the UV cutoff is the expected ``area law" result in $1+1$D \cite{Chandran_2016}, and this is what is obtained after the double truncation. A coefficient of $\frac 1 3$ to the logarithmic scaling is also an expected\footnote{A coefficient of $\frac 1 3$ is expected in the case of two spatial boundaries (such as in the configuration of Fig. \ref{fig:sjdiamond}). The case of one spatial boundary, where a coefficient of $\frac 1 6$  is expected, was also studied in  \cite{Duffy_2022} and agreement with a logarithmic scaling and the $\frac 1 6$ coefficient was confirmed.} universal constant \cite{Calabrese_2009} that the causal set results agree with.

These results in the causal diamond were extended to the massive scalar field theory in \cite{keseman}. In the massive theory, because mass is an intermediate scale lying between the UV (discreteness scale) and IR (diamond size) scales, the same double truncation procedure of the massless theory can be used. This is extremely useful as we lack analytic results in the massive theory to otherwise give us some guidance towards the nature of the eigenvalues. In  \cite{keseman} scalings of the entanglement entropy with both the UV cutoff and mass were studied, and in both cases the expected result of logarithmic scaling with a coefficient of $\frac 1 3$ was obtained. In the same work, the entanglement entropy results were also extended to R\'enyi entropies \cite{Renyi}. The  R\'enyi entropy of order $q$, in terms of the quantities we are working with in our formulation, is given by \cite{Sorkin_2018, keseman}

\begin{equation}
    S^{(q)}=\frac{-1}{1-q}\sum_{\lambda}\ln(\lambda^q-(\lambda-1)^q).
    \label{renyi}
\end{equation}

The solutions to \eqref{geneig} come in pairs of $\lambda$ and $1-\lambda$ and $|\lambda|\geq 1$. Each term in the sum  \eqref{renyi} represents the contribution from one such pair. Similarly, other measures of entropy such as Tsallis entropy \cite{Tsallis:1987eu} can also be calculated and studied in this manner.   

\subsection{Disjoint Causal Diamond regions}

\begin{figure}
    \centering
    \includegraphics[width=.6\linewidth]{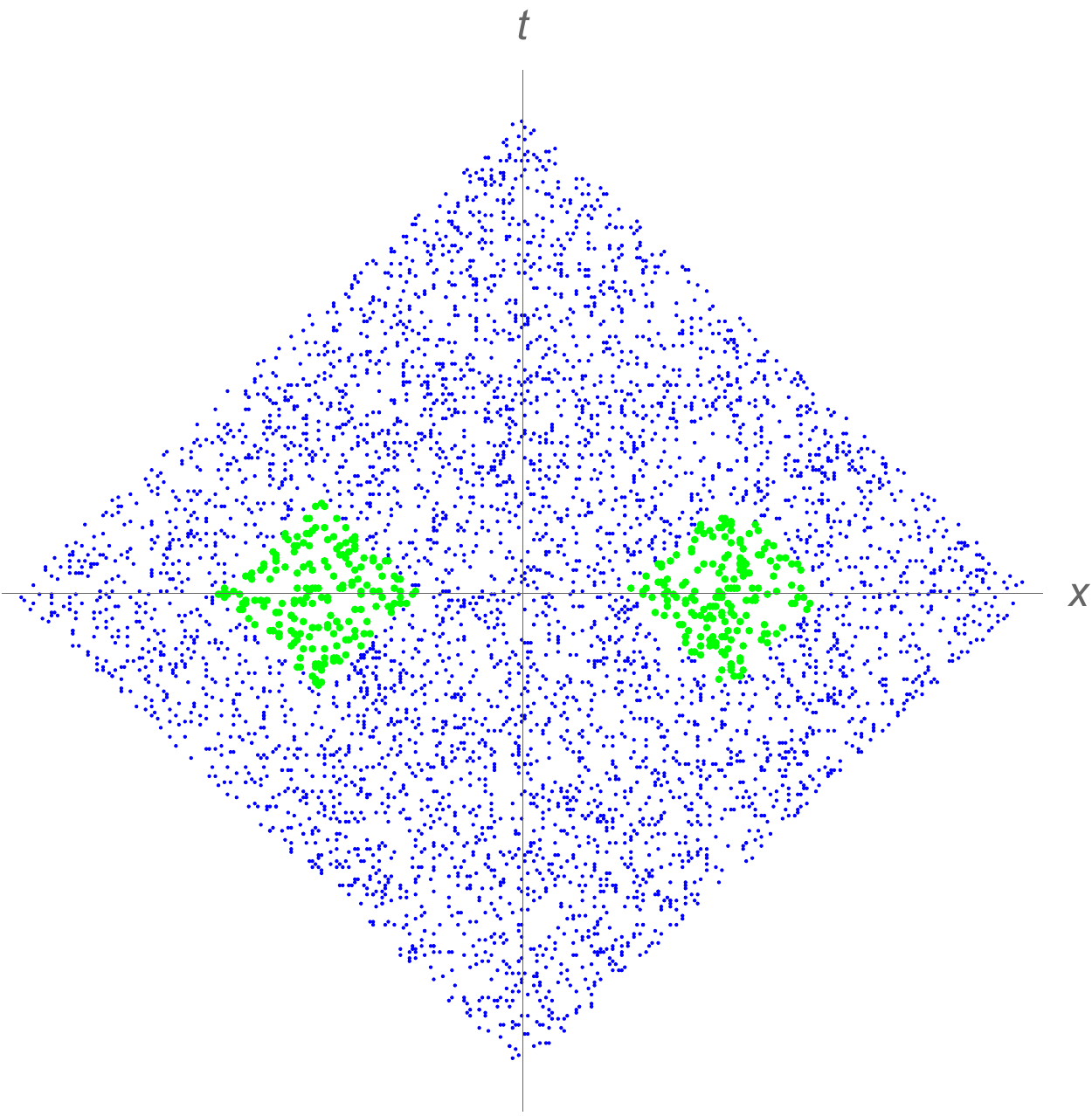}
    \caption{Two smaller causal diamonds within a larger causal diamond.}
    \label{fig:twoints}
\end{figure}

Another setting in which entanglement entropy in causal set theory has been studied, is that of disjoint causal diamonds within a larger global causal diamond in $1+1$D \cite{Duffy_2022}. An example setup with two disjoint subdiamonds is shown in Fig. \ref{fig:twoints}. Entanglement entropy of disjoint regions has been studied in several places in the literature \cite{PhysRevB.81.060411, Ryu_2006, Arias_2018} and the calculations tend to be quite involved and difficult. In contrast, the calculation using \eqref{geneig} and \eqref{s4} for the disjoint diamonds is very similar to the calculation for the single diamond, which is now well-understood and relatively easy to do. Therefore, this is an example where there are calculational advantages, in addition to physical ones, to working with 
the spacetime formulation of Section \ref{subsec:entropy} in a causal set.

In \cite{Duffy_2022} explicit calculations were done for the case of two and three disjoint subdiamonds; the entanglement entropy scalings in both of these cases were shown to be consistent with the logarithmic scalings expected. While scalings with respect to the UV cutoff were the focus of  \cite{Duffy_2022}, there are several other scales in the problem (e.g. the sizes of the subdiamonds, the separation(s) between the diamonds, the distance away from the boundary of the global diamond, etc.) whose relation to the entanglement entropy  would be interesting to investigate.

The mutual information for a two-subdiamond setup was also studied in \cite{Duffy_2022}. The mutual information in this case is the difference between the entanglement entropy associated with the union of the two subdiamonds and the sum of the entropies of the individual subdiamonds. Specifically, the relation between the mutual information and the separation distance between  two diamonds was studied. The results demonstrated the expected qualitative behavior that the mutual information asymptotically vanishes as the separation between the subdiamonds grows and diverges as the separation goes to zero.
\subsection{De Sitter Spacetime}
Cosmological event horizons, just like black hole event horizons,  also have a classical entropy, the Gibbons-Hawking entropy \cite{gh}, associated with them that scales as their spatial area. Hence applications of entanglement entropy to cosmological spacetimes are of particular interest. De Sitter spacetime offers an especially convenient setting to study the entanglement entropy. This is partly due to its maximal symmetry, which makes sprinkling into it considerably easier in comparison to sprinkling into more generic curved spacetimes. Of course, any sprinkling into de Sitter spacetime would not represent the full global de Sitter spacetime, as that has infinite volume and would therefore require an infinite number of elements, which is  computationally not feasible.  Instead, sprinklings into finite slabs of global de Sitter spacetime are used, and it is ensured that any results obtained are stable under making the volume of the slab larger and larger. 

\begin{figure}
    \centering
    \includegraphics[width=.5\linewidth]{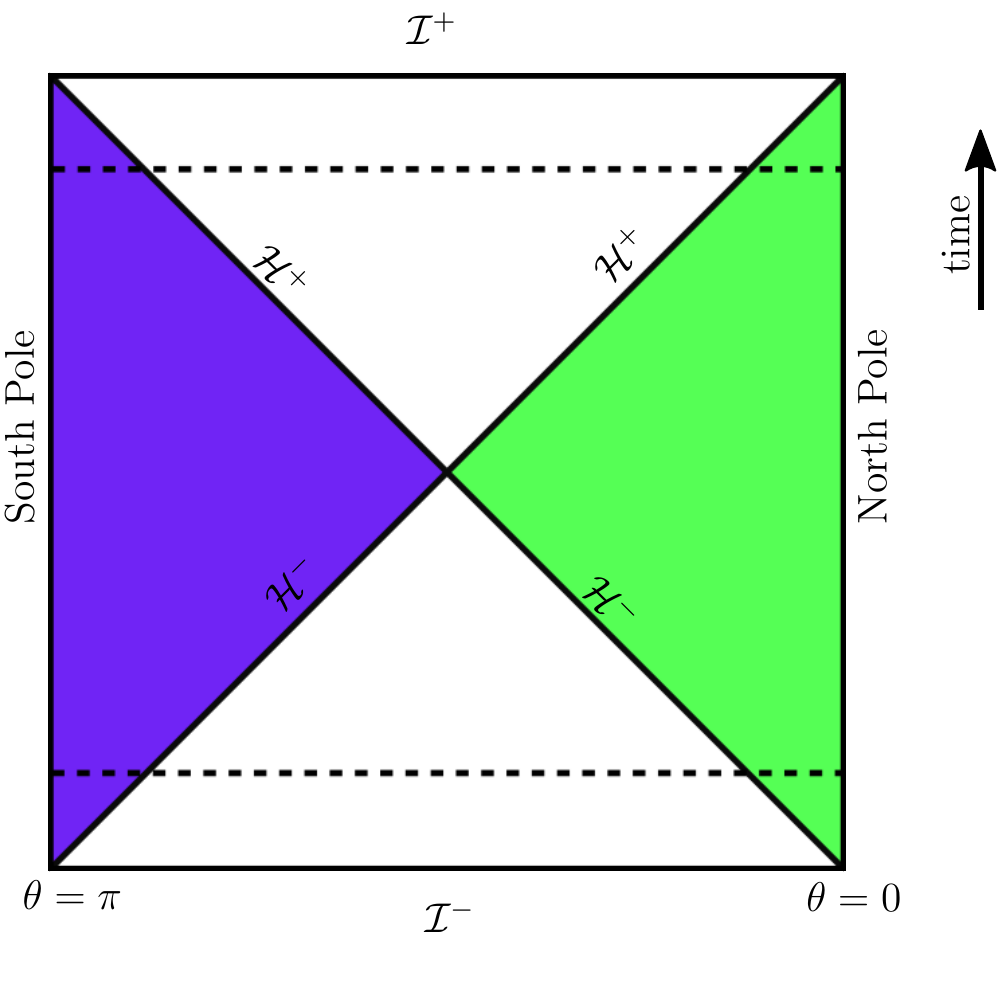}
    \caption{Entangled wedges corresponding to the horizon of a static observer at the North Pole (Green) and its causal complement (Purple). The dashed lines represent the boundaries of the slab, $-T_{max}<T<T_{max}$, in de Sitter spacetime $ds^2=\frac{1}{\cos^2T}\left(-dT^2+d\Omega_{d-1}^2\right)$.}
    \label{fig:dswedge}
\end{figure}

Another attractive feature of working with de Sitter spacetime is that an expression for the retarded Green function, in terms of causal set quantities, is known in this context \cite{nx}. This gives us the starting point in \eqref{scheme}. The SJ Wightman function in causal sets sprinkled into de Sitter spacetime was studied in \cite{sjds}. The entanglement entropy in causal set sprinklings of de Sitter slabs, using the formalism reviewed in this chapter, was studied in \cite{eeds}. In particular the entanglement entropy associated to the subregion within the horizon of a static observer at the North pole was considered. This subregion and its causal complement are shown in the conformal diagram in Fig. \ref{fig:dswedge}.

Similar to the case of the causal diamond, the spectrum of $i\Delta$ has two characteristic regimes: a branch of eigenvalues that are larger in magnitude and follow a power law, and a branch of more numerous small but nonzero eigenvalues that do not follow a power law. Without truncating away  this second branch, once again a spacetime volume scaling is obtained. A spatial area scaling is recovered only after implementing a double truncation. However, choosing a precise double truncation scheme is more subtle in this case, as we lack analytic results in de Sitter spacetime to guide us. In other words, we do not know exactly how the eigenvalues of $i\Delta$ relate to something like a $k_{max}$ in de Sitter space. In the absence of analytic results to guide us, one can estimate the transition between the power law regime and non-power law regime using a number of different strategies. Some of these strategies were studied in \cite{eeds} and shown to yield spatial area laws. However, in that work it was found challenging to hone in on a unique prescription for the truncations, as many different choices yielded area laws. On the other hand, complementarity of the entanglement entropy was found to be a property that was difficult to preserve, since in the second truncation it was unclear which (non-unique) truncation in a subregion ought to be paired with which (non-unique) counterpart truncation in the causally complementary subregion. We will return to this point of finding a general truncation scheme in the concluding section of this chapter.

\subsection{Nonlocal Quantum Field Theory}
\label{subsec:nonlocal}
Ordinarily, the retarded Green function which is the starting point of \eqref{scheme} would be obtained via the Klein Gordon equation in the continuum. In causal set theory, we do not have a local field equation that is the analogue of the Klein Gordon equation. Instead, we must obtain $G_R$ through other means. There is no general recipe in causal set theory for deriving $G_R$. Expressions for it are known in a few distinct cases with help from  the continuum analogues of these Green function and/or dimensional analysis.

While a local analogue of the d'Alembertian $\Box$ and therefore the Klein Gordon equation does not exist in causal set theory, a \emph{nonlocal} analogue of it does \cite{sorkin_box, Benincasa_2010, Dowker_2013, Aslanbeigi_2014}. In fact a whole family $\Box_k$ of nonlocal d'Alembertians exists, with each member distinguished by a nonlocality scale $\ell_k$.
For example in $1+1$D, $\Box_k\phi$ at the element $x\in\mathcal C$ is defined to be \cite{sorkin_box}
\begin{equation}
\Box_k\phi(x)=\frac{4\epsilon}{\ell_\rho^2}\left(\frac{1}{2}\phi(x)+\epsilon\sum_{y\prec x} f\left(n(x,y),\epsilon\right)\phi(y)\right),
\label{box2}
\end{equation}
where $\ell_\rho$ is the discreteness scale,  $\epsilon\equiv\ell_\rho^2\ell_k$,  $n(x,y)$ is the number of elements in the causal diamond between $x$ and $y$, and
\begin{equation}
f(n,\epsilon)=(1-\epsilon)^n\left(1-\frac{2\epsilon n}{1-\epsilon}+\frac{\epsilon^2 n (n-1)}{2 (1-\epsilon)^2}\right).
\end{equation}

The nonlocality of \eqref{box2} is evident in the fact that in order to know the action of the d'Alembertian on the field at the point $x$, we must consider a sum of quantities involving a set of other elements $y$ to the past of $x$. Therefore this nonlocality has a causal, or more specifically retarded, nature.

 In the infinite density limit ($\ell_\rho\rightarrow 0$), the mean of $\Box_k$ over all sprinklings into a spacetime reduces to the usual continuum d'Alembertian plus a
term containing the Ricci scalar curvature \cite{Benincasa_2010}:

\begin{equation}
    \displaystyle{\lim_{\ell_\rho\rightarrow 0}}\, \bar{\Box}_k\, \phi(x)=(\Box-\frac{1}{2}R(x))\, \phi(x).
\end{equation}

With these nonlocal retarded causal set d'Alembertians at hand, we can now invert them to obtain their corresponding retarded Green functions $G_{R,k}$, for use in \eqref{scheme}. This was done in \cite{Belenchia_2018} for nested causal diamonds in $1+1$, $2+1$, and $3+1$-dimensional Minkowski spacetime. Once again, as in the local calculations, only after the use a double truncation, the expected spatial area scalings were obtained.

\section{Discussion and Outlook}
\label{sec:end}
Entanglement entropy in causal set theory is a powerful way to covariantly and unambiguously count the quantum field degrees of freedom one has access to in settings such as spacetimes with event horizons. One must know the scalar field retarded Green function in the spacetime of interest in order to carry out the entanglement entropy calculations, as well as a double truncation rule in order to project out the irrelevant degrees of freedom in the causal set.

An expression for the retarded Green function in causal sets approximated by generic curved spacetimes, in terms of quantities intrinsic to the causal set, is not at present known. Such an expression is known in a few cases, such as the Minkowski and de Sitter examples reviewed above. Nonlocal versions of these Green functions, as discussed in Section \ref{subsec:nonlocal}, can be computed more generally. Alternatively, viewing the causal set as a Lorentzian and covariant discretization of the continuum, we can also simply take the Green functions and/or correlator expressions from the continuum and restrict them to the causal set elements. Thereby we would be regulating any coincidence limit divergences that may be present, by imposing the minimum distance set by the causal set discreteness scale. 

As mentioned, it is also necessary to have a prescription for the double truncation in order to meaningfully study entanglement entropy in causal set theory. This prescription is well understood in the massless theory in causal diamonds in $1+1$D flat spacetime. As shown in \cite{Duffy_2022} and \cite{keseman}, the same prescription can be used for the case of multiple disjoint causal diamonds as well as the massive scalar field theory. More generally, the same prescription can be used in any $1+1$D scalar field theory (e.g. the nonlocal theory or in curved spacetimes), as long as the intermediate scale is far from the discreteness scale. This is because the truncations concern the deep UV regime of the theory, which has the same character in all theories where the other length scales are far from the discreteness scale.

In \cite{eeds} some generalizations of the $1+1$D causal diamond double truncation scheme were studied and applied to calculations in causal diamonds in $3+1$D Minkowski spacetime as well as slabs in de Sitter spacetime. One strategy was to keep $\alpha N^{\frac{D-1}{D}}$ of the largest eigenvalues and their corresponding eigenfunctions, where $\alpha$ is a constant that is a free parameter (several choices for $\alpha$ were investigated). This counting is motivated by the expectation that the number of independent degrees of freedom are the number that would lie within some approximate Cauchy surface-like submanifold (e.g. a thickened antichain). The number of elements in such a submanfiold would be proportional to its volume, which is approximately $V^{\frac{D-1}{D}}$. This strategy succeeds in yielding area laws, but it does not produce a unique prescription (many choices of $\alpha$ are possible) and complementarity is difficult to achieve. Another strategy was to try to estimate the transition between the power law to non-power law regime in the spectrum, through estimating when the approximate linear trend in Fig. \ref{fig:eigs} ends and begins to curve. This strategy sometimes succeeds in producing an area law but it too suffers from an ambiguity in how and how sensitively to define the transition from power law (line on the log-log scale) to non-power law (curve on the log-log scale).
There are some other possible truncation schemes that would merit future investigation. For example, one ansatz could be that in the power law regime, each $n^{th}$ (positive) eigenvalue $\tilde{\lambda}_n$ of $i\Delta$ (in any dimension), when sorted from largest to smallest, is proportional to $\frac {1} {n^p}$, where the proportionality constant is given unambiguously by the value of the largest eigenvalue (where $n=1$) and $p$ can be approximated from the spectrum. For example we know that $p=1$ in the causal diamond in $1+1$D and $p=\frac 1 2$ in the causal diamond in $2+1$D. We can then choose $n_{max}$ to be $N^{\frac{D-1}{D}}$ and estimate the magnitude of the smallest  eigenvalue in the power law regime to then be $\tilde{\lambda}_{min}=\frac{\tilde{\lambda}_{max}}{N^{\frac{p(D-1)}{D}}}$.

More analytic results for the eigenvalues of $i\Delta$ in the continuum would also aid our understanding of the eigenvalues in the causal set and better inform our truncation schemes. It is, however, quite difficult to analytically solve for the eigenfunctions of integral operators.

Another perspective is that there should be no truncations, and that all nonzero eigenvalues and eigenfunctions must contribute to the entropy \cite{Mathur:2022ivs}. However, even in this case we must face the question of how small of an eigenvalue we can really expect to exist in the causal set calculations. Remember that we have the condition \eqref{kernel} that $i\Delta v\neq 0$. Due to the numerical nature of the calculations, there is always some degree of numerical error, and we must identify a threshold beyond which to set the values to zero. While doing so, we must also assess whether we are throwing away anything physical due to the numerical error. Therefore,  a better understanding of the truncated contributions and what solutions can be meaningfully supported on the causal set is needed. 

In \cite{keseman} some insight was gained into the nature of the truncated contributions. Motivated by the observation that the untruncated contributions had continuumlike analogues whereas the truncated ones did not, as well as the observation that the truncated eigenfunctions had many sharp variations at the discreteness scale, it was investigated whether the truncated contributions may be fluctuations particular to a given sprinkling. Namely, it was investigated whether these contributions were random fluctuations that behaved differently from one sprinkling to the next, or whether they had features which persisted over an ensemble of different sprinklings. There are different prescriptions one can use to investigate this; one particular scheme, involving fixing a coarser sub-causal set in order to use it to take averages, was used in \cite{keseman}. Indeed, evidence was found in favor of this conjecture that the truncated contributions are fluctuation-like, and the scheme studied in \cite{keseman} indicated a transition point to the fluctuation-like regime that was consistent with the double truncation in the causal diamond in $1+1$D. This is promising both as insight into the nature of the truncated contributions, as well as practically in order to use it to inform a double truncation scheme in more general settings. The transition to fluctuation-like behaviour can thus potentially be used in general to  distinguish between contributions we must keep and ones we must not. 

There are many other applications of the entanglement entropy formulation reviewed in this chapter that would be interesting to explore in causal set theory. For example, up to first order in perturbation theory, the entanglement entropy for interacting scalar field theories such as those introduced in \cite{Sorkin_2011, emma, kasia}, can be studied. There is also an analogue of \eqref{geneig} for Fermionic field theories, except with the anti-commutator appearing instead of the commutator. Currently, there is no known construction of a Fermionic field theory on a causal set, in terms of quantities intrinsic to the causal set. When such a construction is available, the entanglement entropies of Fermionic field theories could also be studied. 

\bibliographystyle{ieeetr}
 \bibliography{bibliography}{}

\end{document}